\documentclass[conference]{IEEEtran}
\IEEEoverridecommandlockouts
\usepackage{cite}
\usepackage{amsmath,amssymb,amsfonts}
\usepackage{algorithmic}
\usepackage{graphicx}
\usepackage{textcomp}
\usepackage{xcolor}
\usepackage{bm}
\usepackage{float}
\usepackage{booktabs}
\usepackage{url}
\def\BibTeX{{\rm B\kern-.05em{\sc i\kern-.025em b}\kern-.08em
    T\kern-.1667em\lower.7ex\hbox{E}\kern-.125emX}}
\begin{document}

\title{Spectral Codecs: Improving Non-Autoregressive Speech Synthesis with Spectrogram-Based Audio Codecs}

\author{
\IEEEauthorblockN{Ryan Langman, Ante Jukić, Kunal Dhawan, Nithin Rao Koluguri, Jason Li}
\IEEEauthorblockA{\textit{NVIDIA}}
}

\maketitle

\begin{abstract}
Historically, most speech models in machine-learning have used the mel-spectrogram as a speech representation. Recently, discrete audio tokens produced by neural audio codecs have become a popular alternate speech representation for speech synthesis tasks such as text-to-speech (TTS). However, the data distribution produced by such codecs is too complex for some TTS models to predict, typically requiring large autoregressive models to get good quality. Most existing audio codecs use Residual Vector Quantization (RVQ) to compress and reconstruct the time-domain audio signal. We propose a spectral codec which uses Finite Scalar Quantization (FSQ) to compress the mel-spectrogram and reconstruct the time-domain audio signal. A study of objective audio quality metrics and subjective listening tests suggests that our spectral codec has comparable perceptual quality to equivalent audio codecs. We show that FSQ, and the use of spectral speech representations, can both improve the performance of parallel TTS models.
\end{abstract}

\begin{IEEEkeywords}
neural audio codec, text-to-speech, speech synthesis
\end{IEEEkeywords}

\begin{figure*}[t]
  \centering
  \noindent{\includegraphics[width=0.8\paperwidth]{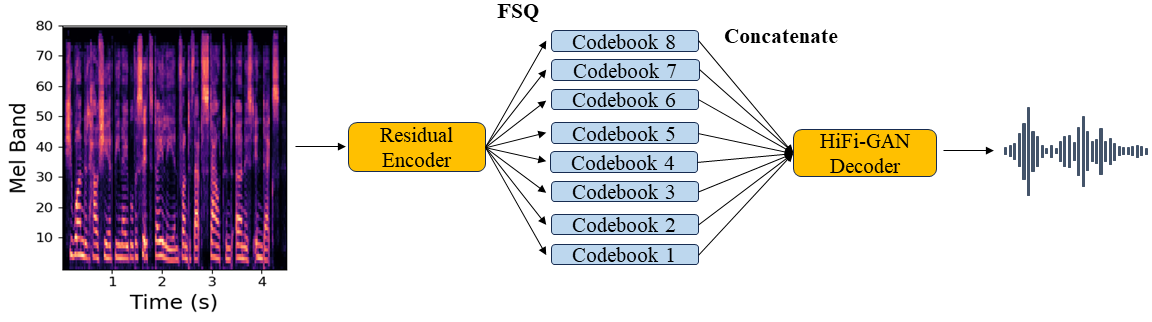}}
  \caption{Architecture of our spectral codec. The 80-dimensional mel-spectrogram is encoded, and then quantized using FSQ. All codebook embeddings are concatenated and given as input to a \mbox{HiFi-GAN} decoder which predicts the final waveform.}
  \label{fig:architecture_mel}
\end{figure*}

\section{Introduction}
\label{sec:intro}

Neural text-to-speech (TTS) systems typically consist of two neural models: an acoustic model which takes text as input and predicts a speech representation such as the mel-spectrogram, and a vocoder which takes the speech representation as input and predicts the audio waveform. Common acoustic models include Tacotron~2~\cite{shen2018natural}, FastPitch~\cite{lancuki2021fastpitch}, and FastSpeech~2~\cite{ren2022fastspeech}. Common vocoders include HiFi-GAN~\cite{kong2020hifigan}, BigVGAN~\cite{lee2023bigvgan}, and UnivNet~\cite{jang2021univnet}.

Neural audio codecs, which encode audio into a discrete latent space, are becoming a popular alternative to spectrogram-based vocoders. Some prominent audio codecs include EnCodec~\cite{défossez2022high}, SoundStream~\cite{zeghidour2021soundstream}, AudioDec~\cite{Wu_2023}, and Descript Audio Codec~\cite{kumar2023highfidelity}. It has been shown that using these audio codecs, architectures for large language models can be adapted for speech synthesis, such as \mbox{VALL-E} \cite{wang2023neural} and SoundStorm~\cite{borsos2023soundstorm}. Though, compared to previous speech synthesis models, these new models are known to hallucinate~\cite{t5tts}, rely on large datasets that are difficult to collect at high sampling rates, and are either autoregressive, or require complex and slow prediction functions such as iterative decoding and delay patterns~\cite{copet2024simple}. These factors make it challenging to utilize such models for many real-world real-time applications.

Audio codecs typically use an encoder-quantizer-decoder network to encode the audio waveform, quantize it, and decode the quantized features to reconstruct the original audio signal. Audio codecs, such as EnCodec, most commonly use a residual vector quantizer (RVQ)~\cite{gray1982} for quantization. 
We explore the use of finite scalar quantization (FSQ)~\cite{mentzer2023finite} for audio codecs, which has been shown to have comparable performance to RVQ in computer vision tasks. The differences between RVQ and FSQ are explained in depth in the methodology section.

In this work, we study the performance of different audio codecs when trained with an autoregressive (AR) TTS model, and a non-autoregressive (NAR) TTS model. We use the term \textit{parallel} in two different ways. First, to refer to parallel models, which are NAR models that predict all outputs at all timesteps in a single iteration. Second, to refer to the ability of a model, either AR or NAR, to predict all codebooks within a timestep in a single iteration.

Our contributions are as follows.

\begin{itemize}
    \item We show NAR TTS performance can be improved by using a spectral codec instead of an audio codec. To the best of our knowledge, this is the first work to demonstrate success predicting discrete speech codes with a parallel model.
    \item To the best of our knowledge, this is the first work to apply FSQ to neural audio codecs. We find that compared to RVQ, FSQ improves TTS performance when predicting codebooks in parallel.
    \item We show that using FSQ does not significantly degrade the quality of the codec reconstruction, when evaluated on speech.
\end{itemize}

Additionally, related work suggests that our spectral codec performs better than existing state-of-the-art audio codecs when trained with autoregressive language models ~\cite{t5tts}.

We publicly release our training code, and pretrained checkpoints, for our audio FSQ codec and spectral codecs in NeMo ~\cite{Harper_NeMo_a_toolkit}. We provide a demo page, with links to individual checkpoints, and audio samples highlighting our results.\footnote{https://rlangman.github.io/spectral-codec/}

\section{Methodology}
\label{sec:method}

In this work we first compare the reconstruction quality of two audio codecs, two spectral codecs, and a HiFi-GAN~\cite{kong2020hifigan} baseline. We then evaluate the audio quality of FastPitch~\cite{lancuki2021fastpitch} trained with each of these six models. For a subset of codecs, we train with an autoregressive version of FastPitch. In this section, we summarize the architectures and training objectives of all models.

All codec models in this study are similar in size, have token rates of 86.1 tokens per second, and bitrates of 6.9\,kbps.

\subsection{Quantization Strategies}
\label{ssec:quantization}

For training our codecs we use two quantization strategies: RVQ and FSQ. RVQ uses a recursive algorithm to create a hierarchical codebook structure. The first codebook contains a high dimensional embedding, encoding most of the content of the audio. Other codebooks encode residuals, which are added to the embedding produced by the first codebook. The value of each residual codebook is determined recursively based on the sum of the first codebook and all previous residual codebooks. We use an RVQ setup similar to EnCodec~\cite{défossez2022high}, with 8 codebooks, 1024 codes per codebook, and a codebook embedding dimension of 128.

FSQ~\cite{mentzer2023finite} takes a continuous embedding and limits the embedding values to the range [-1, 1] using a tanh activation. A number of discrete values is selected for each embedding dimension. For example, if an embedding dimension is configured to have five discrete values, then the tanh output is rounded to the nearest of [-1, -0.5, 0.0, 0.5, 1.0]. For eight discrete values, it is rounded to the nearest of [-0.75, -0.5, -0.25, 0.0, 0.25, 0.5, 0.75, 1.0]. A group of four embedding dimensions configured with [8, 5, 5, 5] discrete values, has a combined total of 1,000 possible discrete values. This can be used to create a l0-bit codebook with 1,000 possible entries.

We follow this FSQ setup for our codecs. The final layer of the codec encoder is projected down to 32 hidden dimensions, followed by a tanh activation. 8 of these dimensions are configured to contain 8 discrete values each, and the other 24 dimensions are configured to contain 5 discrete values each. We group them as mentioned previously, to create 8 codebooks, with 1000 codes per codebook, and a codebook embedding dimension of 32.

\subsection{Spectral Codec Architecture}
\label{ssec:spectralarch}

Our spectral codec is based on the \mbox{HiFi-GAN} architecture, which is state of the art for mel-spectrogram inversion. We use the same mel-spectrogram configuration and decoder architecture as 44.1kHz HiFi-GAN v1. We make the model larger by increasing the decoder initial channels from 512 to 1024. We introduce a residual encoder which encodes and quantizes the mel-spectrogram. The encoder output is projected to the appropriate codebook embedding dimension (32 for FSQ or 128 for RVQ), and then quantized. Instead of the decoder receiving the mel-spectrogram as input, it instead receives the quantized embeddings produced by the encoder. The total size of this model is 65M parameters, with 10M in the encoder and 55M in the decoder.

\begin{table*}[t]
  \caption{Speech reconstruction quality for signals reconstructed using codecs with different input features and quantizers.}
  \label{table:reconstruction}
  \centering
  \resizebox{\textwidth}{!}
  {
  \begin{tabular}{c c | c c c c c c}
    \toprule
    Input Feature & Quantizer & MOS $\uparrow$ & ViSQOL $\uparrow$ & Mel Distance $\downarrow$ & STFT Distance $\downarrow$ & SI-SDR $\uparrow$ & SI-SDR (Squim) $\uparrow$ \\
    \midrule
    ground truth & -- & 3.96~$\pm~0.06$ & -- & -- & -- & -- & -- \\
    audio waveform & RVQ & 3.75~$\pm~0.06$ & 4.29~$\pm~0.01$ & 0.115~$\pm~0.001$ & 0.034~$\pm~0.000$ & $\textbf{7.87}~\boldsymbol{\pm}~\textbf{0.14}$ & 21.65~$\pm~0.20$ \\
    audio waveform & FSQ & $\textbf{3.89}~\boldsymbol{\pm}~\textbf{0.06}$ & 4.28~$\pm~0.01$ & 0.114~$\pm~0.001$ & 0.034~$\pm~0.000$ & 7.76~$\pm~0.14$ & $\textbf{21.72}~\boldsymbol{\pm}~\textbf{0.20}$ \\
    mel-spectrogram & -- & 3.71~$\pm~0.06$ & $\textbf{4.70}~\boldsymbol{\pm}~\textbf{0.00}$ & $\textbf{0.071}~\boldsymbol{\pm}~\textbf{0.001}$ & $\textbf{0.031}~\boldsymbol{\pm}~\textbf{0.000}$ & -23.08~$\pm~0.35$ & 18.32~$\pm~0.19$ \\
    mel-spectrogram & RVQ & $\textbf{3.89}~\boldsymbol{\pm}~\textbf{0.06}$ & $\textbf{4.42}~\boldsymbol{\pm}~\textbf{0.01}$ & $\textbf{0.102}~\boldsymbol{\pm}~\textbf{0.001}$ & 0.034~$\pm~0.000$ & -23.62~$\pm~0.32$ & 19.02~$\pm~0.19$ \\
    mel-spectrogram & FSQ & 3.80~$\pm~0.06$ & 4.37~$\pm~0.01$ & 0.109~$\pm~0.001$ & 0.035~$\pm~0.000$ & -22.85~$\pm~0.33$ & 19.45~$\pm~0.19$ \\
    \bottomrule
  \end{tabular}
  }
\end{table*}

\begin{table*}[t]
  \caption{Speech synthesis quality for TTS models trained with different codecs.\\RTF values are reported relative to the mel-spectrogram baseline without quantization.\\We omit MOS for TTS models which produce poor audio quality based on metrics and informal listening tests.}
  \label{table:tts}
  \centering
  \begin{tabular}{c c c | c c c c c c}
    \toprule
    Input Feature & Quantizer & Model & MOS $\uparrow$ & ViSQOL $\uparrow$ & ESTOI $\uparrow$ & WER / \% $\downarrow$ & CER / \% $\downarrow$ & RTF $\downarrow$ \\
    \midrule
    ground truth & -- & -- & 4.35~$\pm~0.04$  & -- & -- & 2.25~$\pm~0.48$ & 0.44~$\pm~0.09$ & -- \\
    audio waveform & RVQ & Parallel & -- & 3.14~$\pm~0.03$ & 0.60~$\pm~0.01$ & 3.13~$\pm~0.62$ & 1.01~$\pm~0.23$ & 0.80 \\
    audio waveform & FSQ & Parallel & -- & 3.07~$\pm~0.02$ & 0.56~$\pm~0.01$ & 2.56 ~$\pm~0.53$ & 0.68~$\pm~0.15$ & 0.80 \\
    mel-spectrogram & -- & Parallel & 3.45~$\pm~0.06$ & 3.86~$\pm~0.03$ & 0.74~$\pm~0.01$ & 2.52~$\pm~0.52$ & 0.68~$\pm~0.18$ & 1.00 \\
    mel-spectrogram & RVQ & Parallel  & 3.75~$\pm~0.05$ & 3.57~$\pm~0.03$ & 0.69$\pm~0.01$ & 2.36~$\pm~0.49$ & 0.55~$\pm~0.10$ & 0.85 \\
    mel-spectrogram & FSQ & Parallel & 3.63~$\pm~0.06$ & $\textbf{3.84}~\boldsymbol{\pm}~\textbf{0.03}$ & $\textbf{0.77}~\boldsymbol{\pm}~\textbf{0.01}$ & \textbf{2.30}\boldsymbol{~$\pm~0.47$} & \textbf{0.50}\boldsymbol{~$\pm~0.10$} & 0.87 \\
    audio waveform & RVQ & Autoregressive & 3.39~$\pm~0.07$ & 2.75~$\pm~0.06$ & 0.56~$\pm~0.01$ & 19.54~$\pm~2.69$ & 15.11~$\pm~2.22$ & 15.63 \\
    audio waveform & FSQ & Autoregressive & $\textbf{4.15}~\boldsymbol{\pm}~\textbf{0.05}$ & 3.68~$\pm~0.03$ & $\textbf{0.77}~\boldsymbol{\pm}~\textbf{0.01}$ & 2.34~$\pm~0.49$ & 0.53~$\pm~0.12$ & 13.41 \\
    mel-spectrogram & FSQ & Autoregressive & 3.67~$\pm~0.06$ & 3.80~$\pm~0.03$ & 0.76~$\pm~0.01$ & \textbf{2.30}\boldsymbol{~$\pm~0.49$} & 0.51~$\pm~0.11$ & 13.07 \\
    \bottomrule
  \end{tabular}
\end{table*}

\subsection{Audio Codec Architecture}
\label{ssec:audioarch}

To create a symmetric audio codec, we use the same \mbox{HiFi-GAN} V1 decoder architecture, with 768 initial channels, and upsample rates $[8, 8, 4, 2]$. We create an encoder by inverting this decoder, using the same inversion scheme as EnCodec~\cite{défossez2022high}. The encoder has 48 initial channels and downsampling rates of $[2, 4, 8, 8]$. The total model size is 62M parameters, with 31M in the encoder and 31M in the decoder. We train one version of the audio codec with an RVQ and one with FSQ.

\subsection{Codec Training Objective}
\label{ssec:trainobj}

For reconstruction loss we use the multi-resolution mel-spectrogram loss from \cite{kumar2023highfidelity}, and the multi-resolution log-magnitude short-time Fourier transform (STFT) loss introduced in~\cite{jang2021univnet}, with window lengths [32, 64, 128, 256, 512, 1024, 2048] and 25\% hop length. The mel-spectogram loss uses mel dimensions [5, 10, 20, 40, 80, 160, 320]. We use the multi-period discriminator introduced in \cite{kong2020hifigan} and the multi-scale complex STFT discriminator introduced in \cite{défossez2022high} both with squared-GAN and feature matching loss. As in~\cite{défossez2022high} we update the discriminators only once every two steps. All training losses have a weight of 1.0, except the STFT loss which has a weight of 20.0.

\subsection{TTS Architecture}
\label{ssec:ttsarch}

For our TTS architecture we use FastPitch~\cite{lancuki2021fastpitch}. FastPitch is a non-autoregressive transformer-based encoder-decoder system which takes text, pitch, and duration as input to predict a mel-spectrogram. To improve quality we also condition the model on energy as in~\cite{ren2022fastspeech} and extract duration information during training using the alignment strategy in~\cite{badlani2021tts}.

FastPitch predicts continuous mel-spectrogram features with an L2 loss. To predict audio tokens, we replace the L2 loss with a softmax loss predicting codebooks. All codecs have eight codebooks, resulting in eight softmax layers which predict the entries of the different codebooks in parallel. For the softmax loss we include logit normalization~\cite{wei2022mitigating} with a temperature $\tau=0.02$.

For some experiments we introduce an AR decoder to FastPitch. The AR decoder is the same architecture as the original NAR decoder, except with causal attention heads, and linear feed forward networks. At training time the ground truth audio codes are shifted by one time step, passed through two 512 dimension linear projections, and added to the AR decoder input. During inference the decoder input at each time step is conditioned on the code prediction for the previous time step.

Our FastPitch encoder and decoder each consist of 6 transformer blocks with 512 latent dimension and 2048 dimensional feed forward networks. The size of the NAR model is 88M parameters, and the size of the AR model is 108M parameters.

\section{Experiments}
\label{sec:exp}

\subsection{Datasets}
\label{ssec:data}

While most speech domains use 16\,kHz audio, many TTS applications require 44.1\,kHz data. For TTS training we use HiFi-TTS~\cite{bakhturina2021hifi}, containing 300 hours of high quality 44.1\,kHz audio from 10 speakers.

To train a robust audio codec we need a larger dataset. However most public datasets that are uploaded at 44.1\,kHz contain upsampled mixed-bandwidth data. When audio codecs are trained on mixed bandwidth data, their quality degrades significantly as they fail to reconstruct frequencies above some bandwidth threshold~\cite{puvvada2024discrete}. In \cite{kumar2023highfidelity} the authors suggest that the threshold is the average bandwidth seen at training time.

To address this, we create a large-scale 44.1\,kHz dataset containing only full-bandwidth data. We take the metadata from the English subset of MLS~\cite{Pratap_2020} and download the original 48\,kHz audiobooks from the Librivox website. We downsample to 44.1\,kHz and use the bandwidth estimation strategy described in HiFi-TTS~\cite{bakhturina2021hifi} to filter out all data with an estimated bandwidth below 16\,kHz. In this way we create a new dataset containing 12.8k hours of full-bandwidth 44.1\,kHz speech with 3.1M utterances and 2,776 speakers. We use this dataset for training our audio codecs. For evaluation we use the same test set as MLS with approximately 3.8k utterances from 42 speakers not seen in the training data.

\subsection{Training}
\label{ssec:train}

All models were trained using NVIDIA NeMo~\cite{Harper_NeMo_a_toolkit} with eight V100 GPUs.

All codec models were trained for 100k training steps with a batch size of 16 examples per GPU and 16,384 audio samples ($\approx{0.37}$ seconds) per example, using an Adam optimizer with learning rate $2 \cdot 10^{-4}$, $\beta_1=0.8$, $\beta_2=0.99$, and exponential learning rate decay with $\gamma=0.998$ per 1,000 steps.

All FastPitch models were trained for 1M training steps with a batch size of 8 per GPU, using an AdamW optimizer with learning rate $5 \cdot 10^{-3}$, weight decay $10^{-6}$, $\beta_1=0.9$, $\beta_2=0.999$, and a Noam learning rate schedule with 1,000 warmup step. 

\section{Results}
\label{sec:result}

\subsection{Codec Evaluation}
\label{ssec:codeceval}

We conduct an ablation study on the reconstruction quality of the codec models. The results are in Table~\ref{table:reconstruction}. As a baseline, we also include a mel decoder with no quantization which is equivalent to HiFi-GAN.

To evaluate performance we use a combination of instrumental metrics and subjective listening tests. To evaluate perceptual quality we use MOS and ViSQOL~\cite{hines2015visqol}. For measuring time-domain accuracy we use SI-SDR~\cite{roux2018sdr}. For measuring spectral accuracy we look at the L1 distance between log mel-spectrogram and log magnitude STFT features using window length 2048 and hop length 512.

For MOS evaluation, listeners are provided with one audio sample at a time and asked to rate the audio quality on scale from 1 (bad quality, very noisy) to 5 (very clear, no noise). 100 randomly selected test utterances are evaluated by 15 listeners.

\subsection{Codec Performance}
\label{ssec:codecperf}

The MOS study shows that the perceived quality between most codec models is not statistically significant, with listeners showing a small preference for the ground truth, and a small preference against the audio RVQ codec and HiFi-GAN.

The instrumental metrics between all of the codecs are comparable, with the spectral codecs performing better on spectral accuracy, and the audio codecs performing better on time-domain accuracy. Alongside the MOS study, these results suggests that the input feature and quantization approach used have minimal impact on the perceptual quality of the reconstructed speech.

Interestingly, we see that our spectral codecs have poor SI-SDR. This is because time-domain metrics, such as SI-SDR, assume that waveforms are aligned. The spectral codec estimates phase from the spectrogram, resulting in audio which sounds realistic but is not aligned with the ground truth. When we use Squim~\cite{kumar2023torchaudiosquim} to estimate SDR without a reference, the results are more closely correlated with the perceptual quality metrics.

\subsection{TTS Evaluation}
\label{ssec:ttseval}

We conduct an ablation study on the TTS quality of FastPitch when trained to predict the tokens of the five codec models, and the continuous mel-spectrogram with the HiFi-GAN baseline. The results are in Table~\ref{table:tts}.

To evaluate perceptual quality we use MOS and ViSQOL~\cite{hines2015visqol}. To measure intelligibility of TTS outputs we look at ESTOI~\cite{7539284}, as well as the word error rate (WER) and character error rate (CER) of transcriptions using NVIDIA's Fast Conformer-Transducer XL~\cite{fastconformer_transducer_xlarge}. We include the Real-Time Factor (RTF) for each TTS system, normalized by the RTF of our HiFi-GAN baseline. All metrics are reported with a $95\%$ confidence interval. For WER and CER we compute the confidence intervals with bootstrapping \cite{1326009}.

\subsection{TTS Performance}
\label{ssec:ttsperf}

The TTS models trained with spectral codecs are significantly less noisy than with the HiFi-GAN baseline. This suggests that quantization in the vocoder could be a solution to the issue of over-smoothness~\cite{ren2022revisiting} that is common when using continuous speech representations.

Metrics and listening tests suggest that the TTS models trained with RVQ codecs have significant audio quality issues. This is consistent with past works that find that predicting an RVQ codebook requires conditioning on codebooks earlier in its hierarchy \cite{wang2023neural}\cite{borsos2023soundstorm}\cite{t5tts}. We see that TTS performance improves when using FSQ. The AR model predicting RVQ codebooks has particularly poor performance, and sometimes produces unintelligible speech producing high WER and CER. In comparison the AR model trained with the audio FSQ codec produces high quality speech, as indicated by the MOS results. This demonstrates that FSQ codebooks can be effectively predicted in parallel within a timestep, which is consistent with the results of \cite{t5tts}.

The performance of the NAR TTS model is significantly better across metrics when using spectral codecs compared to audio codecs. This suggests that spectral features can be predicted more accurately in parallel across different timesteps.

Interestingly, we do observe that the AR model performs better with the audio FSQ codec than with the spectral codecs. This seems inconsistent with the results of \cite{t5tts} that showed improvements in AR performance when using a spectral codec. An initial analysis of the MOS data suggests that FastPitch performs worse on speakers in HiFi-TTS whose speech data contains diverse acoustic conditions. This might indicate that ground truth information other than pitch, such as context or reference audio, is needed to learn TTS on such datasets. However, we leave further analysis and potential solutions for future work.

\section{Conclusion}
\label{sec:conc}

In this work we explore the performance of AR and NAR TTS models with different audio codecs. We observe that AR TTS models can synthesize high quality speech with standard audio codecs, but that predicting audio tokens in parallel is challenging. We find that NAR models have better performance with spectral codecs than they do with audio codecs. We also find that TTS models, that predict codebooks within a timestep in parallel, have better performance with codecs trained with FSQ than with codecs trained with RVQ. Furthermore, we show that codecs trained with FSQ have comparable reconstruction quality to those trained with RVQ.

\clearpage

\bibliographystyle{IEEEtran}
\bibliography{refs}

\end{document}